\documentclass[aps,prc,twocolumn,showpacs,amssymb,superscriptaddress]{revtex4-1}
\date{\today}

\usepackage{graphicx}% Include figure files
\usepackage{dcolumn}% Align table columns on decimal point
\usepackage{float}
\usepackage{subfigure}
\usepackage[tbtags]{amsmath}
\usepackage{xcolor}
\usepackage{xspace}
\usepackage{braket}
\usepackage{hyperref}
\usepackage{booktabs}

\newcolumntype{d}{D{.}{.}{-1}}

\AtBeginDocument{
\heavyrulewidth=.08em
\lightrulewidth=.05em
\cmidrulewidth=.03em
\belowrulesep=.65ex
\belowbottomsep=0pt
\aboverulesep=.4ex
\abovetopsep=0pt
\cmidrulesep=\doublerulesep
\cmidrulekern=.5em
\defaultaddspace=.5em
}

\newcommand{\bb}{$\beta\beta$\xspace}
\newcommand{\bbz}{$0\nu\beta\beta$\xspace}

\begin{document}

\title{Neutrinoless double-beta decay matrix elements in large shell-model
spaces with the generator-coordinate method}

\author{C.F.\ Jiao}
\affiliation{Department of Physics and Astronomy, University of North Carolina, Chapel Hill, North Carolina, 27599-3255,
USA}\affiliation{Department of Physics, Central Michigan University, Mount Pleasant, Michigan, 48859,
USA}

\author{J. Engel}
\email[Email: ]{engelj@physics.unc.edu}
\affiliation{Department of Physics and Astronomy, University of North Carolina,
Chapel Hill, North Carolina, 27599-3255, USA}

\author{J.D.\ Holt}
\email[Email: ]{jholt@triumf.ca}
\affiliation{TRIUMF, 4004 Wesbrook Mall, Vancouver, BC V6T 2A3, Canada}

\begin{abstract}
We use the generator-coordinate method with realistic shell-model interactions
to closely approximate full shell-model calculations of the matrix elements for
the neutrinoless double-beta decay of $^{48}$Ca, $^{76}$Ge, and $^{82}$Se.  We
work in one major shell for the first isotope, in the $f_{5/2}pg_{9/2}$ space
for the second and third, and finally in two major shells for all three.  Our
coordinates include not only the usual axial deformation parameter $\beta$, but
also the triaxiality angle $\gamma$ and neutron-proton pairing amplitudes.  In
the smaller model spaces our matrix elements agree well with those of full
shell-model diagonalization, suggesting that our Hamiltonian-based GCM captures
most of the important valence-space correlations.  In two major shells, where
exact diagonalization is not currently possible, our matrix elements are only
slightly different from those in a single shell. 
\end{abstract}


\maketitle

\section{\label{intro}Introduction}

Experiments to measure the rate of neutrinoless double-beta (\bbz) decay are
increasing in number and scale \cite{avi08}, in part because they offer the only
real hope of determining whether neutrinos are Majorana particles.  The rate of
decay, however, depends on nuclear matrix elements that must be accurately
calculated to allow experimentalists to plan efficiently and interpret results.
At present, the predictions of various nuclear models for the matrix elements
differ by factors of up to three~\cite{engel16,vog12}, and it is not possible to
estimate the theoretical uncertainty in any of them.  Improving the accuracy of
matrix-element calculations has become an important goal for the
nuclear-structure community.

The methods used to calculate the matrix elements include the shell
model~\cite{Iwata16, sen16, men08}, the interacting boson model
(IBM)~\cite{bar15}, the quasiparticle random phase approximation (QRPA)
~\cite{Mustonen13, sim13, Hyvarinen15}, and the generator coordinate method
(GCM)~\cite{rod10, vaq13, yao15, hin14}.  Both the QRPA and the GCM have been
used in conjunction with energy density functional (EDF)
theory~\cite{Mustonen13, vaq13, yao15}. These methods, which allow large
single-particle spaces (and thus allow unrestricted collective deformation and
pairing) but do not contain all kinds of correlations, yield matrix elements
that are usually larger than those of the shell model, which employs only a few
single-particle levels but allows arbitrarily complex correlations within them.
Recent work to compare the shell-model with the GCM \cite{men14} suggests that
the extra valence-space correlations, particularly those due to isoscalar
pairing \cite{men16}, are responsible for much of the difference in the
predictions of the two methods.  

In an attempt to include all relevant physics, Ref.~\cite{hin14} proposed
combining the virtues of the large-single-particle-space methods with those of
the shell model by using the neutron-proton pairing amplitudes as generator
coordinates in the GCM approach, within spaces of two or more major shells.
Though promising, the results in that paper were obtained with the restriction
(as in prior GCM work on \bbz decay \cite{rod10,vaq13,yao15}) that deformation
be axially symmetric and with a simple pairing-plus-quadrupole interaction.
Here we include a measure of triaxial deformation parameter as a generator
coordinate and work with genuine shell-model interactions.  We use a one-shell
calculation to compare our results, obtained with carefully fit and widely used
interactions \cite{men08}, to those of exact diagonalization with the same
interactions.  Having verified the accuracy of our approach, we move on to a
two-shell spaces, in which exact diagonalization is not currently possible.
Because no careful Hamiltonian tuning has yet been done for the $fp-sdg$ model
space appropriate for $A\approx 80$, we use the perturbation-theoretic methods
discussed in Ref.\ \cite{Tsunoda14} to derive an interaction. Our results are
the first for $^{76}$Ge and $^{82}$Se of a realistic shell-model-like
calculation in more than one shell, and the first exploration of the effects of
triaxiality on $\beta\beta$ decay.  Recent experiments indicate that the
deformation of $^{76}$Ge is indeed triaxial \cite{Toh13}.

This paper is organized as follows: Section~\ref{mode} presents a brief overview
of the \bbz matrix elements and of the GCM with a Hamiltonian.
Section~\ref{calc1} contains results in a single shell, which we use to test our
approach by comparing with exact diagonalization.  Section \ref{calc2}, which
presents \bbz matrix elements in two shells, is the heart of the paper.
Section~\ref{summary} contains a preview of future work and a summary.

\section{\label{mode}Matrix Elements and Methods}

In the closure approximation, the quantity we need is the matrix element of a
two-body operator between the ground states of the initial and final nuclei.  If
the decay is produced by the exchange of a light-Majorana neutrino with the
usual left-handed currents, 
%and if in Eq.\ \eqref{eqn:ME} we neglect the tensor
%term (which is small for the decay of $^{76}$Ge and $^{82}$Se according to
%Refs.\ \cite{men08} and \cite{kor07a}), 
we can write the nuclear matrix element as \cite{pan96,sim08}
\begin{widetext}
\begin{align}
\label{eqn:ME}
M^{0\nu}&=M^{0\nu}_{\text{GT}}-\frac{g^2_V}{g^2_A}M^{0\nu}_{\text{F}}
+ M^{0\nu}_T \\
&= \frac{2R}{\pi g^2_{A}}\int_0^{\infty}\!\! q \, dq \, 
\bra{F} \sum_{a,b}\frac{j_0(qr_{ab})\left[h_{\text{GT}}(q)\vec{\sigma}_a\cdot
\vec{\sigma}_b + h_{\text{F}}(q)\right]+j_2(qr_{ab})h_T(q)\left[3\sigma_1 \cdot
\vec{r}_{ab} \sigma_2 \cdot \vec{r}_{ab} - \sigma_1 \cdot \sigma_2\right]}
{q+\overline{E}-(E_I+E_F)/2}\tau^+_a\tau^+_b\ket{I} \,, \nonumber
\end{align}
\end{widetext}
where GT, F, and T refer to the Gamow-Teller, Fermi, and tensor parts of the
matrix element. The vector and axial coupling constants are given by $g_V=1$ and
$g_A \approx 1.27$, $|I\rangle$ and $|F\rangle$ are the ground-states of the
initial and final nuclei, $r_{ab}$ is the distance between nucleons $a$ and $b$,
$j_0$ and $j_2$ are the usual spherical Bessel functions, $\bar{E}$ is an
average excitation energy (to which the matrix element is not sensitive), and
the nuclear radius $R=1.2A^{1/3}$ fm makes the matrix element dimensionless. The
functions $h_{\text{F}}(q)$, $h_{\text{GT}}(q)$, and $h_\text{T}(q)$ contain
nucleon form factors and forbidden corrections to the weak current.  We modify
our wave functions at short distances with the ``Argonne'' correlation
function~\cite{sim09}.  A detailed presentation of the form of the matrix
element can be found in Ref.\ \cite{sim08}. 

The crucial ingredients in Eq.\ \eqref{eqn:ME} are the initial and final ground
states $\ket{I}$ and $\ket{F}$.  To obtain them, we use a shell-model effective
Hamiltonian $H_{\text{eff}}$ in a valence space whose size we are free to
choose.
%\begin{equation}
%H_{\text{eff}} = \sum_{a}\epsilon_a\hat{n}_a 
%                        + \sum_{a\leqslant b, c\leqslant d}\sum_{JT}V_{JT}(ab; cd)\hat{T}_{JT}(ab; cd),
%\end{equation}
%where $\epsilon_a$ stands for single-paritcle energies, $V_{JT}(ab; cd)$ stands for two-body matrix elements (TBME's), $\hat{n}_a$ is the number operator for the spherical orbit $a$ with quantum numbers $(n_a, l_a, j_a)$ and 
%\begin{equation}
%\hat{T}_{JT}(ab; cd) = \sum_{MT_z}A^{\dagger}_{JMTT_z}(ab)A_{JMTT_z}(cd)
%\end{equation}
%is the scalar two-body density operator for nucleon pairs in orbits $a, b$ and
%$c, d$ coupled to quantum numbers $J, M, T, $ and $T_z$. 
The first step in the GCM procedure is to generate a set of reference
quasiparticle vacuua $\ket{\varphi(q_1,q_2,\ldots)}$ that provide the minimum
energy such states can have while constrained to also have expectation values
$q_i = \braket{\mathcal{O}_i}$ for a set of collective operators
$\mathcal{O}_i$.  Here we take the operators $\mathcal{O}_i$ to be: 
\begin{equation}
\label{eq:params}
\begin{aligned}
\mathcal{O}_1 &= Q_{20}\,,&
\mathcal{O}_2 &= Q_{22} \,,   \\
\mathcal{O}_3 &= \frac{1}{2}(P_0+P_0^\dag) \,, & 
\mathcal{O}_4 &= \frac{1}{2}(S_0+S_0^\dag) \,,
\end{aligned}
\end{equation}
where 
\begin{equation}
\label{eq:constr-ops}
\begin{aligned}
Q_{2M} &= \sum_a r^2_a Y^{2M}_a \,, \\ 
P_0^\dag &= \frac{1}{\sqrt{2}}\sum_l \sqrt{2l+1} \, [c^\dag_l
c^\dag_l]^{L=0,J=1,T=0}_{000}  \,, \\
S_0^\dag &= \frac{1}{\sqrt{2}}\sum_l \sqrt{2l+1} \, [c^\dag_l
c^\dag_l]^{L=0,J=0,T=1}_{000} \,,
\end{aligned}
\end{equation}
with $M$ labeling the angular-momentum $z$-projection and $a$ labeling nucleons,
and the brackets signifying the coupling of orbital angular momentum, spin, and
isospin to various values, each of which has $z$-projection zero.  The operator
$c^\dag_l$ creates a particle in the single-particle level with orbital angular
momentum $l$.  The operator $P_0^\dag$ creates a correlated isoscalar pair, and
the operator $S_0^\dag$ a correlated isovector neutron-proton pair.  We actually
only constrain one of the two pair amplitudes at a time: the isoscalar amplitude
when computing $M^{0\nu}_{GT}$ (and $M^{0\nu}_T$, which is small) and the
isovector amplitude when computing $M^{0\nu}_{F}$.  The usual deformation
parameters $\beta$ and $\gamma$ are related to $q_1 \equiv \braket{Q_{20}}$ and
$q_2 \equiv \braket{Q_{22}}$ by $\beta = (\chi b^2 / \omega_0) \sqrt{q_1^2 + 2
q_2^2}$ (with $b$ the oscillator length, given by $2R/\sqrt{5} \,
(3A/2)^{-1/6}$, $\omega_0 =41.2 A^{-1/3}$, $\chi=0.4$) and $\gamma =
\tan^{-1}(\sqrt{2}q_2/q_1)$.

%where b = (4/5) *(2/3)^(1/3) * r_0^2 A ^(1/3) is harmonica oscillator length. For 76Ge, beta = 0.438* /chi * sqrt( <Q_20>^2 + 2 * <Q_22> ^2 ).      

To efficiently include the effect of neutron-proton pairing, we start, as in
Ref.\ \cite{hin14}, from a Bogoliubov transformation that mixes neutrons and
protons, i.e.\ from quasiparticle operators of the (schematic) form
\begin{equation}
\label{eqn:operator}
\alpha^{\dagger}\sim u_pc^{\dagger}_p+v_pc_p+u_nc^{\dagger}_n+v_nc_n.
\end{equation}
In the full equations single-particle states are summed over, so that each of
the coefficients $u$ and $v$ are replaced by matrices, as described in
Ref.\ \cite{goo79}.  We then solve constrained Hartree-Fock-Bogoliubov (HFB)
equations, minimizing expectation values of the form 
\begin{equation}
\label{eqn:H}
\begin{split}
\braket{H^{\prime}}&=\braket{H_{\text{eff}}}-\lambda_Z \left(\braket{N_Z}-Z
\right) -\lambda_N \left(\braket{N_N} - N\right) \\
&-\sum_i \lambda_i \left(\braket{\mathcal{O}_i}-q_i\right) \,,
\end{split}
\end{equation}
where the $N_Z$ and $N_N$ are the proton and neutron number operators,
$\lambda_Z$ and $\lambda_N$ are corresponding Lagrange multipliers, the sum over
$i$ includes up to three of the 4 $\mathcal{O}_i$ in Eq.\ \eqref{eq:params}, and
the other $\lambda_i$ are Lagrange multipliers to constrain the expectation
values of those operators to $q_i$.  We solve these equations many times,
constraining each time to a different point on a mesh in the space of $q_i$.

Having obtained a set of HFB vacua with various amounts of axial deformation,
triaxial deformation, and isoscalar/isovector pairing, we construct the GCM
state by superposing projected HFB vacua: 
\begin{equation}{\label{eqn:GCMwf}}
\ket{\Psi^{J}_{NZ\sigma}}=\sum_{K,q}f^{JK}_{q\sigma}\ket{JMK;NZ;q} \,,
\end{equation}
where $|JMK;NZ;q\rangle\equiv
\hat{P}^J_{MK}\hat{P}^N\hat{P}^Z|\varphi(q)\rangle$ and $q$ is short for the set
of all $q_i$.  Here, the $\hat{P}^{\prime}$s are projection operators onto
states with well-defined angular momentum $J$ and $z$-component $M$, neutron
number $N$, and proton number $Z$~\cite{ring80}. The weight functions
$f^{JK}_{q\sigma}$, where $\sigma$ enumerates states with the same quantum
numbers, follow from the Hill-Wheeler equations~\cite{ring80}
\begin{equation}
\label{eqn:HWeq}
\sum_{K^{\prime},q^{\prime}}\Bigl\{\mathcal{H}_{KK^{\prime}}^J(q;
q^{\prime})-E^J_{\sigma}\mathcal{N}^J_{KK^{\prime}}(q;
q^{\prime})\Bigr\}f^{JK^{\prime}}_{q^{\prime}\sigma}=0,
\end{equation}
where the Hamiltonian kernel $\mathcal{H}_{KK^{\prime}}^J(q; q^{\prime})$ and the norm kernel $\mathcal{N}_{KK^{\prime}}^J(q; q^{\prime})$ are given by
\begin{equation}
\quad
\begin{aligned}
\mathcal{H}_{KK^{\prime}}^J(q; q^{\prime}) &=
\bra{\varphi(q)}H_{\text{eff}}\hat{P}^{J}_{KK^{\prime}}\hat{P}^{N}\hat{P}^{Z}
\ket{\varphi(q^{\prime})} \,, \\
\mathcal{N}_{KK^{\prime}}^J(q; q^{\prime}) &=
\bra{\varphi(q)}\hat{P}^{J}_{KK^{\prime}}\hat{P}^{N}\hat{P}^{Z}
\ket{\varphi(q^{\prime})} \,.
\end{aligned}
\quad
\end{equation}
To solve Eq.\ \eqref{eqn:HWeq}, we first diagonalize the norm kernel
$\mathcal{N}$ and then use the nonzero eigenvalues and corresponding
eigenvectors to construct a set of ``natural states.'' Finally, we diagonalize
the Hamiltonian in the space of these natural states to obtain the GCM states
$|\Psi^{J}_{NZ\sigma}\rangle$ (for details, see Refs.~\cite{Rodriguez10a,
Yao10}).  We carry out this entire procedure in both the initial and final
nucleus, using the lowest $J=0$ states in each as the ground states between
which we sandwich the \bbz operator to obtain the matrix element $M^{0\nu}$ from
Eq.\ \eqref{eqn:ME}. 

\section{\label{calc1}Tests in a Single Shell}

Before undertaking a two-major-shell calculation, we need to test our GCM with a
realistic interaction in a model space small enough to allow exact
diagonalization.  We begin by performing GCM calculations in the $pf$-shell,
comprising the $0f_{7/2}$, $0f_{5/2}$, $1p_{3/2}$, and $1p_{1/2}$ orbits.  Using
use the KB3G interaction~\cite{Poves01}, which accounts successfully for the
spectroscopy, electromagnetic and Gamow-Teller transitions, and deformation of
$pf$-shell nuclei~\cite{cau05}, we compute the \bbz matrix elements of
$^{48}$Ca, $^{54}$Ti, and $^{54}$Cr. Although the last two nuclei are not
candidates for an experiment, they offer opportunities to test the GCM.  Because
these nuclei show no evidence of triaxial deformation, we need only use the
axial quadrupole moment $q_1 \equiv \braket{Q_{20}}$ and isoscalar pairing
amplitude $\phi\equiv q_3=1/2\braket{P_0+P_0^\dag}$ as generator coordinates for
the computation of $M^{0\nu}_{GT}$. 

\begin{figure}
\centering
\includegraphics[width=\columnwidth]{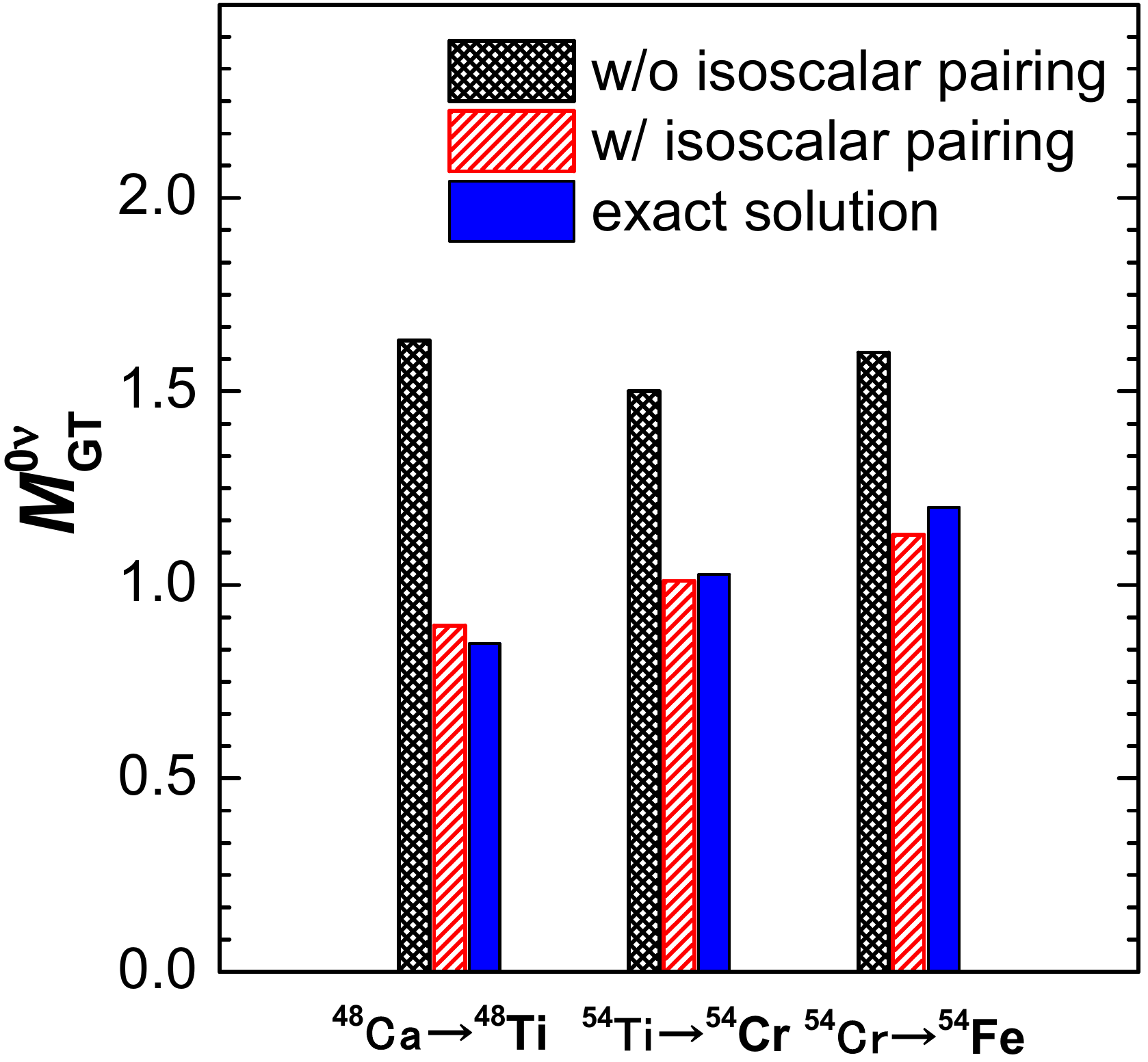}
\caption{\label{fig:caticr}GCM results for the Gamow-Teller part of \bbz matrix
elements of $^{48}$Ca, $^{54}$Ti, and $^{54}$Cr, compared with the results of
exact diagonalization.} 
\end{figure}

\begin{figure}[t]
\begin{minipage}[t]{1\linewidth}
\centering
\includegraphics[width=3.0in]{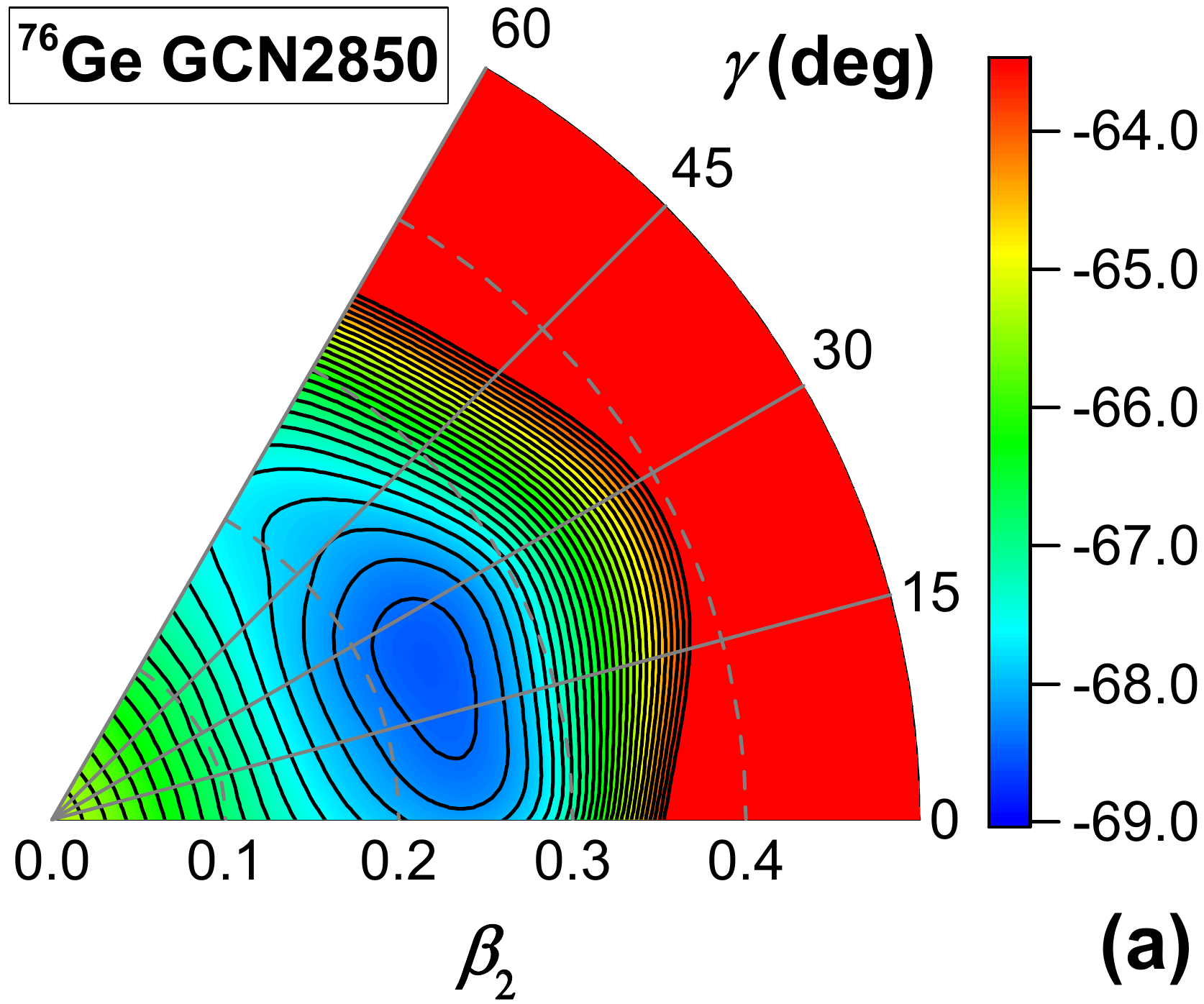}
\label{fig:side:a}
\end{minipage}%
\\
\vspace{0.2in}
\begin{minipage}[t]{1\linewidth}
\centering
\includegraphics[width=3.0in]{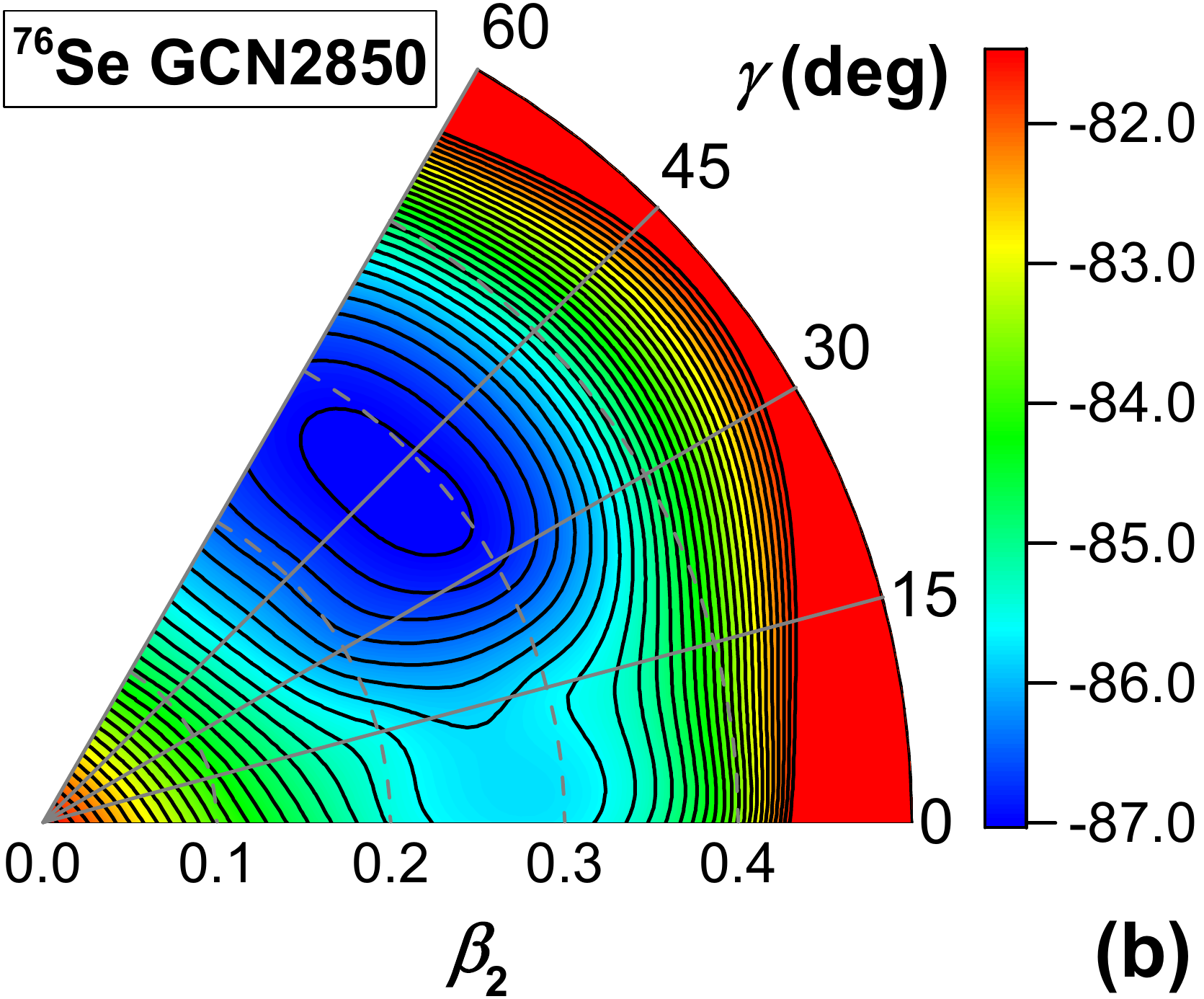}
\label{fig:side:b}
\end{minipage}
\caption{\label{PES} Projected potential-energy surfaces produced by the GCN2850
interaction, with the isoscalar pairing amplitude $\phi=0$, in the
($\beta,\gamma$) plane for $^{76}$Ge and $^{76}$Se. }
\end{figure}

Figure \ref{fig:caticr} shows the GT matrix elements that result from this
procedure, alongside those coming from exact diagonalization.  To highlight the
effects of isoscalar pairing in the GCM, we present the results of two separate
GCM calculations.  In the first, as in Ref.\ \cite{men16}, we set all the
two-body matrix elements of the Hamiltonian with angular momentum $J=1$ and
isospin $T=0$ to zero, because those are the ones through which isoscalar
pairing acts.  The resulting GT matrix elements overestimate the exact one
substantially.  In the second calculation, we use the full KB3G interaction,
with the result that the matrix element decreases, coming quite close to the
exact one.  The sensitivity to isoscalar pairing, pointed out long ago for the
QRPA in Refs.\ \cite{vog86} and \cite{eng88} and more recently for the GCM and
shell model in Refs.\ \cite{hin14} and ~\cite{men16}, shows that the
neutron-proton mixing in our HFB states is essential.  The good agreement with
exact diagonalization suggests that once it is included, we are not omitting
anything of importance.

We turn now to one of the nuclei in which we are really interested: $^{76}$Ge,
used or to be used in many \bb
experiments~\cite{kla01a,Aalseth02,GERDA13,MAJORANA14}.  Shell model
calculations of the \bbz decay of this nucleus
\cite{men08,caurier08,sen16,sen14b} have usually been set in the so-called
$f5pg9$ space, comprising the $0f_{5/2}$, $1p_{3/2}$, $1p_{1/2}$, and $0g_{9/2}$
orbits, and have employed either the JUN45~\cite{Honma09} or
GCN2850~\cite{Gniady} Hamiltonian.  The $f5pg9$ model space is not a complete
major shell; it includes levels from two different major shells and is missing,
in particular, the spin-orbit partners of the $0f_{5/2}$ and $0g_{9/2}$ orbits.
We discuss the effects of including these and other orbits later. 

As we already mentioned, both theory \cite{Guo07,Shen11} and experiment
\cite{Chou93a,Toh13} indicate triaxial deformation in low-lying states of
even-even Ge and Se isotopes near $A=76$. Our calculations predict it as well.
Figure~\ref{PES} displays the $^{76}$Ge and $^{76}$Se quantum-number-projected
potential-energy surfaces (PES's), at isoscalar-pairing amplitude $\phi=0$,
produced by the GCM with the GCN2850 interaction.  The minimum is at
$\beta_2=0.23, \gamma=24^{\circ}$ in $^{76}$Ge, a result that agrees well with
those of EDF-based GCM calculations~\cite{Guo07}, and at at $\beta_2=0.28,
\gamma=45^{\circ}$ in $^{76}$Se.  In addition to this ``static'' triaxial
deformation, dynamical triaxial effects arise from the $\gamma$-soft PES's in
both isotopes.  The GCM, which mixes states with a range of $\gamma$ values,
incorporates dynamical effects.  
%[The energy surfaces at other values of $\phi$ are similar.] 

\begin{figure}[b]
\centering
\includegraphics[width=\columnwidth]{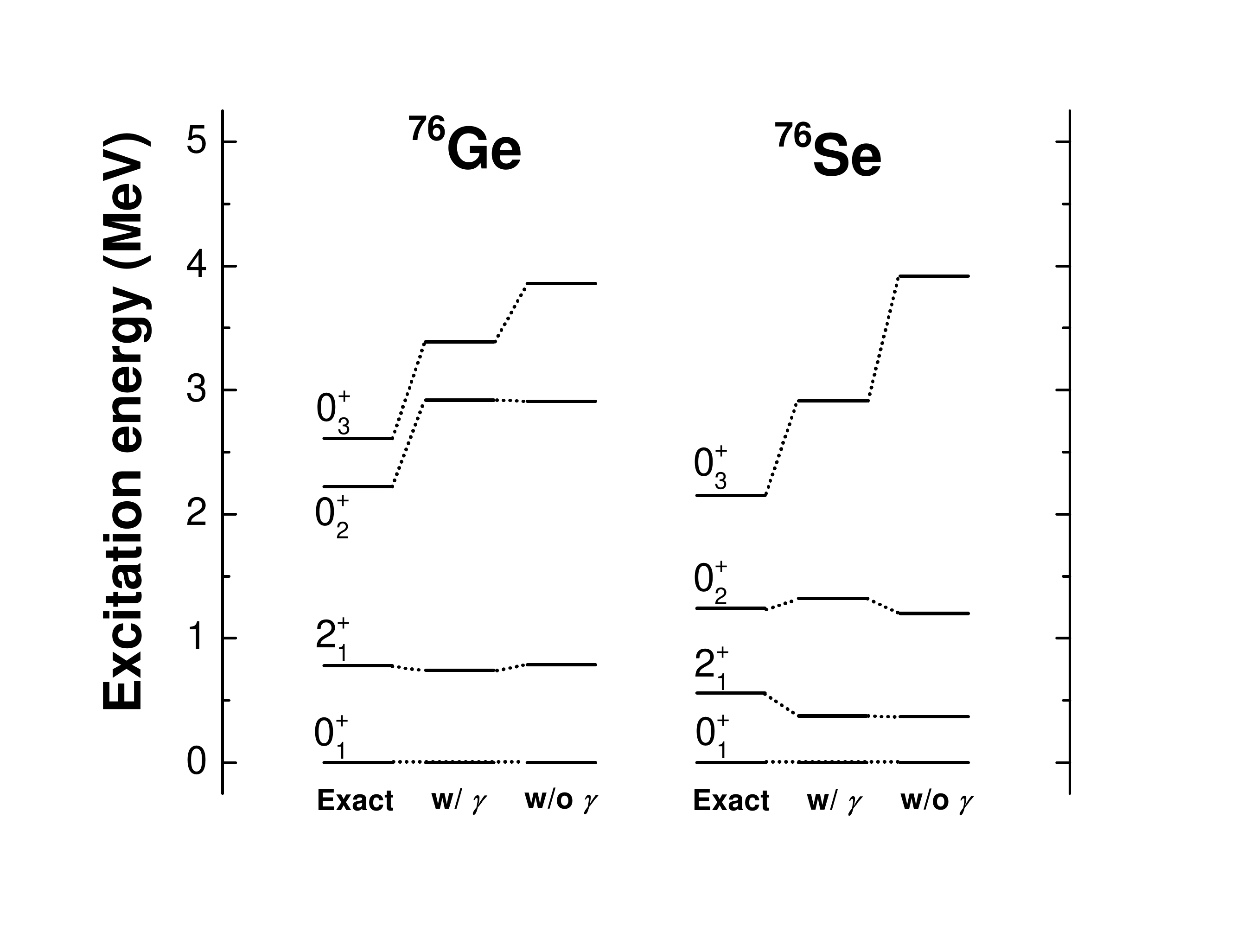}
\caption{\label{fig:def} Low-lying excitation spectra of $^{76}$Ge and $^{76}$Se
produced in one shell by the GCM with the GCN2850 interaction, with and
without triaxial deformation (labeled by the parameter $\gamma$).  The results
from the exact diagonalization of the shell-model Hamiltonian appear for
comparison \cite{men-exact}. }
\end{figure}

Our complete calculations include as generator coordinates both deformation
parameters $q_1$ and $q_2$ (or equivalently $\beta$ and $\gamma$) as well as one
of the proton-neutron-pairing parameters $q_3$ and $q_4$.  We can assess the
effects of triaxial shape fluctuations by including or excluding triaxially
deformed configurations from the set of GCM basis states.  Including them has
clear effects on spectroscopy.  Fig \ref{fig:def} shows the spectra of low-lying
$0^+$ and $2^+$ states in the two important $A=76$ isotopes with GCN2850;
triaxial shapes, though they have a relatively small effect on the first excited
$0^+$ state, lower the second such state significantly in both nuclei, and in
$^{76}$Se by 
%nearly 500 keV,   
over an MeV.  The values for the strength $B(E2; 0^+\rightarrow 2^+)$ are
affected in a similar way.  With triaxial deformation (and with the usual
effective charges $e^\text{eff}_p = 1.5 e$ and $e^\text{eff}_n = 0.5 e$) the
values in $e^2b^2$ are $0.172$ in $^{76}$Ge (vs.\ the exact-diagonalization
value of 0.158) and $0.275$ in $^{76}$Se (vs.\ the exact value of $0.209$).
Without triaxial deformation the numbers are smaller:  $0.154$ in $^{76}$Ge, and
$0.268$ in $^{76}$Se.

Triaxial deformation has a non-negligible effect on the \bbz matrix element as
well.  As Table \ref{t:one-shell} shows, our full GCM calculation gives values
for the matrix elements $M^{0\nu}$ that are about $15\%$ smaller than the
results obtained without triaxially deformed configurations.  The full matrix
elements, though slightly suppressed, are in good agreement with those of exact
diagonalization (in this calculation only, we neglected the very small matrix
element $M^{0\nu}_T$).  The GCM approach with neutron-proton pairing indeed
captures most of correlations around the Fermi surface that are important for
\bbz decay.  The small discrepancy may be due to fluctuations in like-particle
pairing, which we do not treat here but which, according to the EDF-based work
of Ref.\ \cite{vaq13}, increase \bbz matrix elements slightly.  We could include
those fluctuations, but at the cost of a considerable increase in computing
time. 

\begin{table}[t]
\centering
\begin{tabular}{ldd}
\toprule &\multicolumn{1}{c}{GCN2850}&\multicolumn{1}{c}{JUN45} \\ \midrule
Axial GCM    & 2.71                       & 3.42  \\
Triaxial GCM & 2.33                       & 2.94  \\ 
Exact        & 2.81 \ \text{\cite{men08}} & 3.37 \  \text{\cite{sen14b}} \\
\bottomrule 
\end{tabular}
\caption{\label{t:one-shell}The matrix elements $M^{0\nu}$ produced in the GCM
by GCN2850 and JUN45 for the decay of $^{76}$Ge, with and with out triaxial
deformation as a generator coordinate, and by those same interactions with exact
diagonalization.}
\end{table}

\section{Results in Two Shells}
\label{calc2}

The promise of the Hamiltonian-based GCM is an eventual \textit{ab initio}
calculation.  Here we take a step in that direction by working in the full
$fp-sdg$ two shell space.  The number of states for $A=76$ nuclei in this space
is still too large for exact diagonalization.  

Before considering Ge and Se, we make one more test, for $^{48}$Ca, the one
experimental candidate in which an exact two-shell calculation is almost
possible at present.  Ref.\ \cite{Iwata16} uses the SDPFMU-DB interaction, with
the omission of some cross-shell excitations, to compute the \bbz matrix element
nearly exactly.  Our GCM result, 1.082, is close to 1.073, the result of Ref.\
\cite{Iwata16}, and suggests in addition that the cross-shell excitations
neglected in that paper really are unimportant.  With some confidence in the
performance of the GCM in two shells, we turn to the decay of $^{76}$Ge.

The first issue we must grapple in this mid-shell nucleus is what to use for the
valence-space Hamiltonian.  Ref.\ \cite{hin14} used a multi-separable collective
Hamiltonian that we wish to improve on here.  The size of the two-shell space,
however, makes the usual procedure, in which shell-model Hamiltonians are tuned
to data, difficult to follow; furthermore, there are no well-tested Hamiltonians
for this space on the market. The first step in the usual approach is to produce
an initial valence-space Hamiltonian, traditionally in many-body perturbation
theory.  Deficiencies in the many-body method are then remedied by tuning
single-particle energies and interaction matrix elements to experimental data.
Here we must settle for adjusting only single-particle energies.  The tuning of
interaction matrix elements requires repeated calculations that are simply too
time consuming. 

Although nonperturbative methods such as the in-medium similarity
renormalization group can produce shell-model Hamiltonians
\cite{tsukiyama12,bog14}, they have not been tested systematically for valence
spaces larger than one major harmonic-oscillator shell.  We therefore use the
Extended Krenciglowa-Kuo (EKK) variant of many-body perturbation theory
\cite{Tsunoda14}, suitable for non-degenerate valence spaces, to construct an
effective Hamiltonian from a third-order $Q$-box in the $pf-sdg$ shell.  We
begin from the 1.8/2.0 two- plus three-nucleon (3N) interaction of
Refs.~\cite{Hebeler11,Simonis16}; the interaction reproduces ground-state
energies across the light- and medium-mass regions of the nuclear chart
\cite{Simonis17}.  With $\hbar\omega\!=\!10$ MeV, a space of 13 major shells for
intermediate-state sums is enough to ensure convergence.

The monopole components of our valence-space Hamiltonian are particularly
sensitive to the initial three-nucleon interaction \cite{ots10}, which one
generally reduces to effective zero-, one- and two-body parts via normal
ordering with respect to some independent-particle reference state
\cite{hagen07}.  For shell model calculations, the usual reference state is the
inert core, containing all orbitals below the valence space.  As discussed in
Refs.~\cite{Stroberg16,Stroberg17}, however, this choice completely omits
three-nucleon interactions among the valence particles themselves, and if the
target nucleus (here $^{76}$Ge or $^{76}$Se) is far from the core, the neglected
effects become sizable. Since we can choose to normal-order with respect to any
reference we want, it makes sense to include more orbitals, so that we better
capture the bulk effects of three-nucleon interactions among valence particles.
Thus, we take the reference state to be the (fictional) inert core corresponding
to $^{56}$Ni.  While this nucleus is still some distance in proton and neutron
number from the $A=76$ nuclei, we expect it to make a better reference than than
the $^{40}$Ca core.  We call the resulting valence-space interaction $pfsdg$ and
use it exclusively in the following.

\begin{figure}
\begin{minipage}[t]{1\linewidth}
\centering
\includegraphics[width=\textwidth]{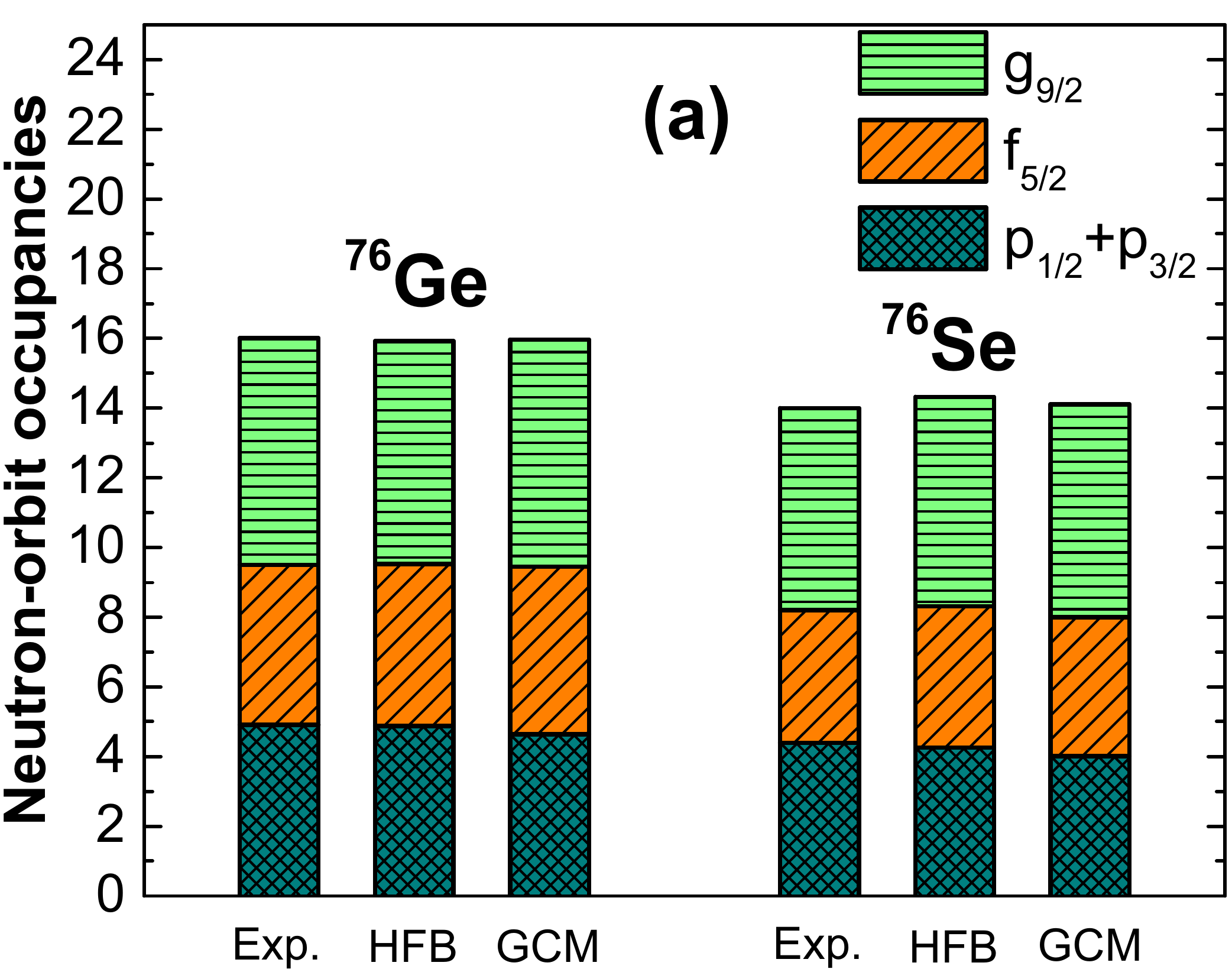}
\end{minipage}%
\\
\begin{minipage}[t]{1\linewidth}
\centering
\includegraphics[width=\textwidth]{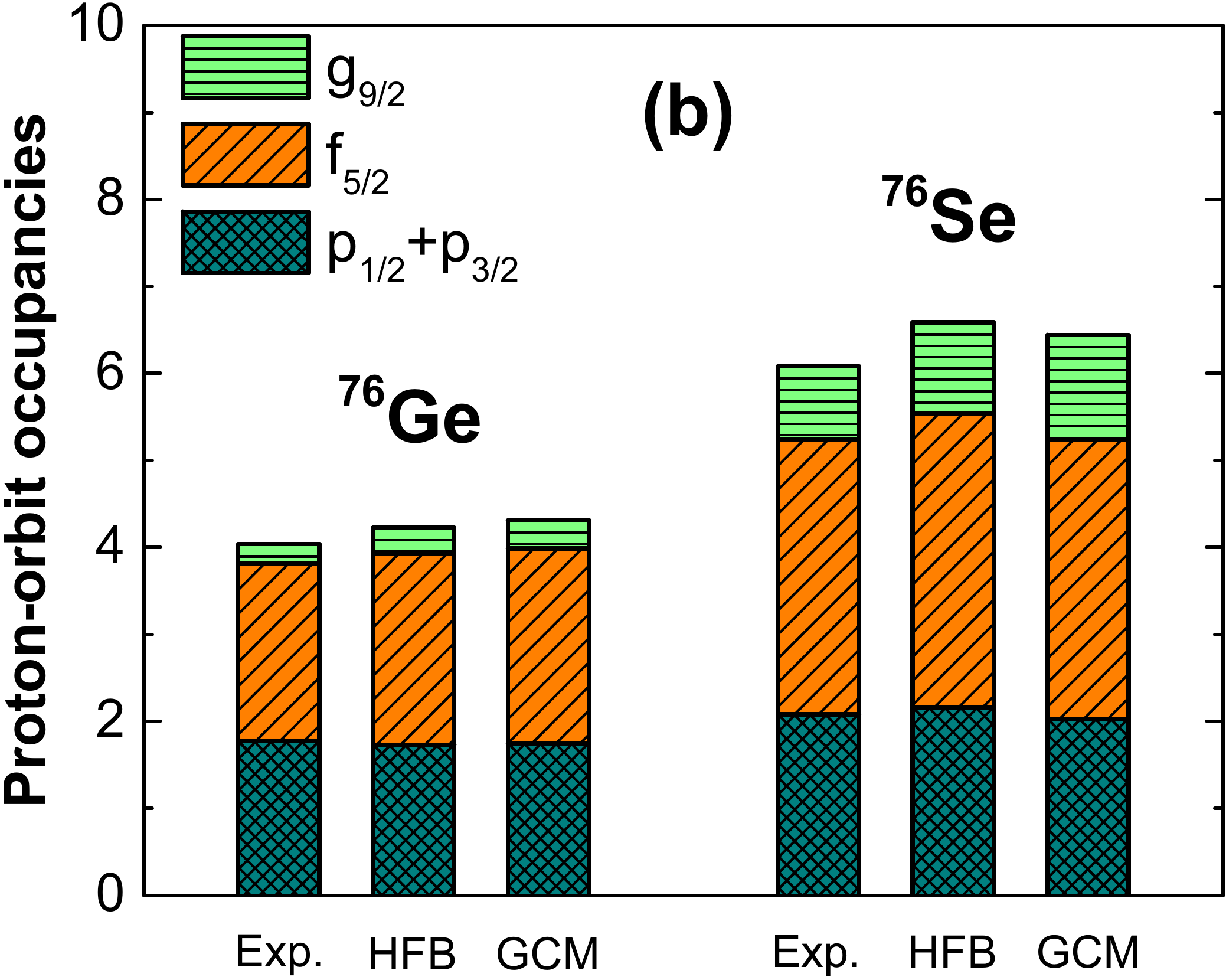}
\end{minipage}
\caption{\label{fig:occ} The occupancies of valence neutron and proton orbits
produced by the interaction $pfsdg$ (see text) for $^{76}$Ge and $^{76}$Se,
following the adjustment of the single-particle energies for levels in the lower
shell.  The measured occupancies are from Refs.~\cite{Schiffer08, Kay09}.} 
\end{figure}

Perhaps because of the non-ideal reference state, the single-particle energies
that emerge from the perturbative procedure are poor. The proton sub-shell gap
at $Z=34$ is too large, causing the proton pairing mean field to disappear no
matter what the deformation and leading to occupation numbers that differ
significantly from measured values.  To remedy the problem, we adjust the
single-particle energies of the orbits in the lower shell to reproduce those
values, while leaving the two-body part of the Hamiltonian untouched.  Figure
\ref{fig:occ} shows the occupation numbers after adjustment, for both the
projected HFB state with the minimum energy (which we used to make the
adjustments) and in the final GCM states.  Though the occupations change a
little when the HFB states are mixed in the GCM, they remain close to the
experimental values.

\begin{figure}[t]
\begin{minipage}{\columnwidth}
\includegraphics[width=\textwidth]{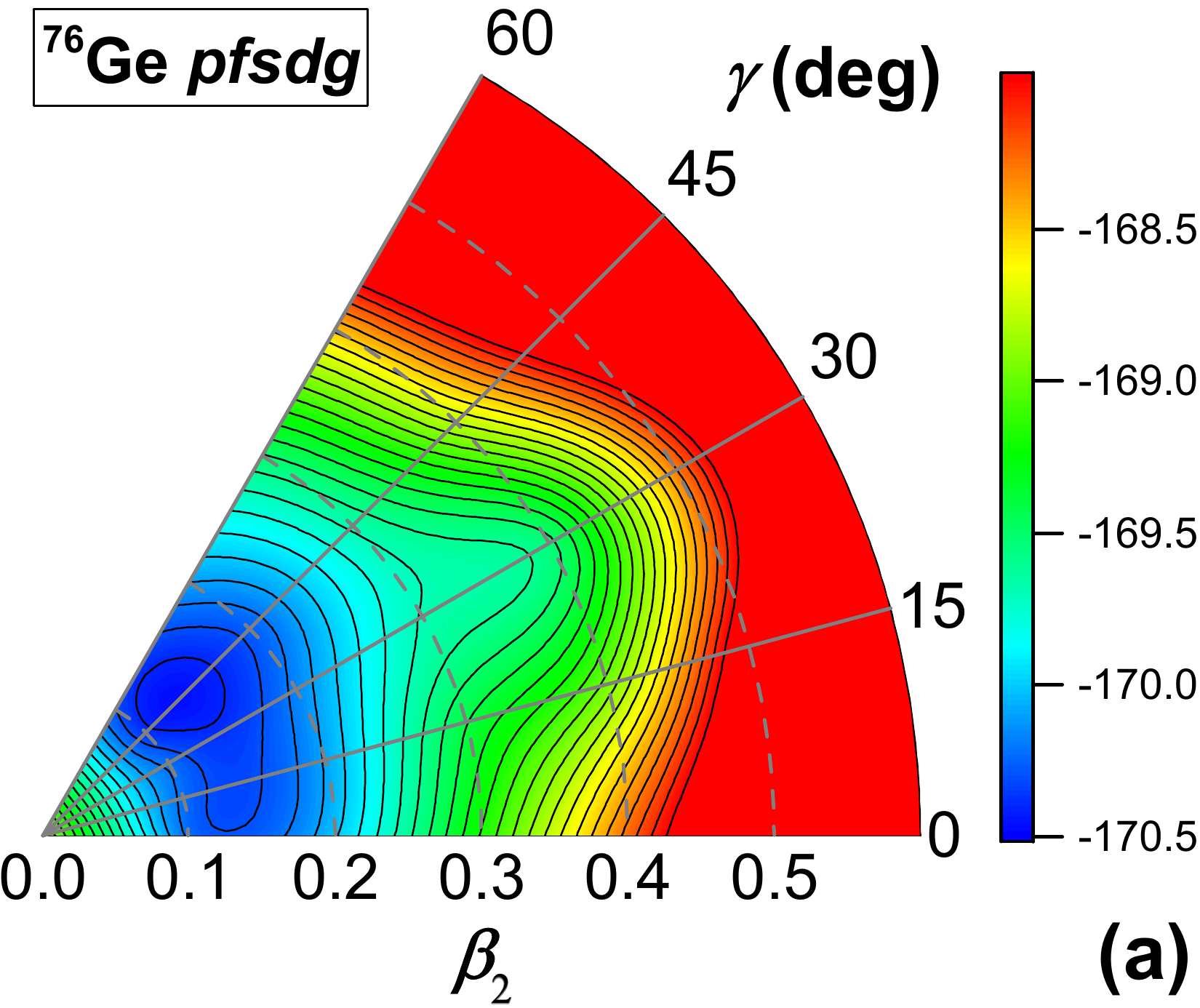}
\end{minipage}\\
\begin{minipage}{\columnwidth}
\includegraphics[width=\textwidth]{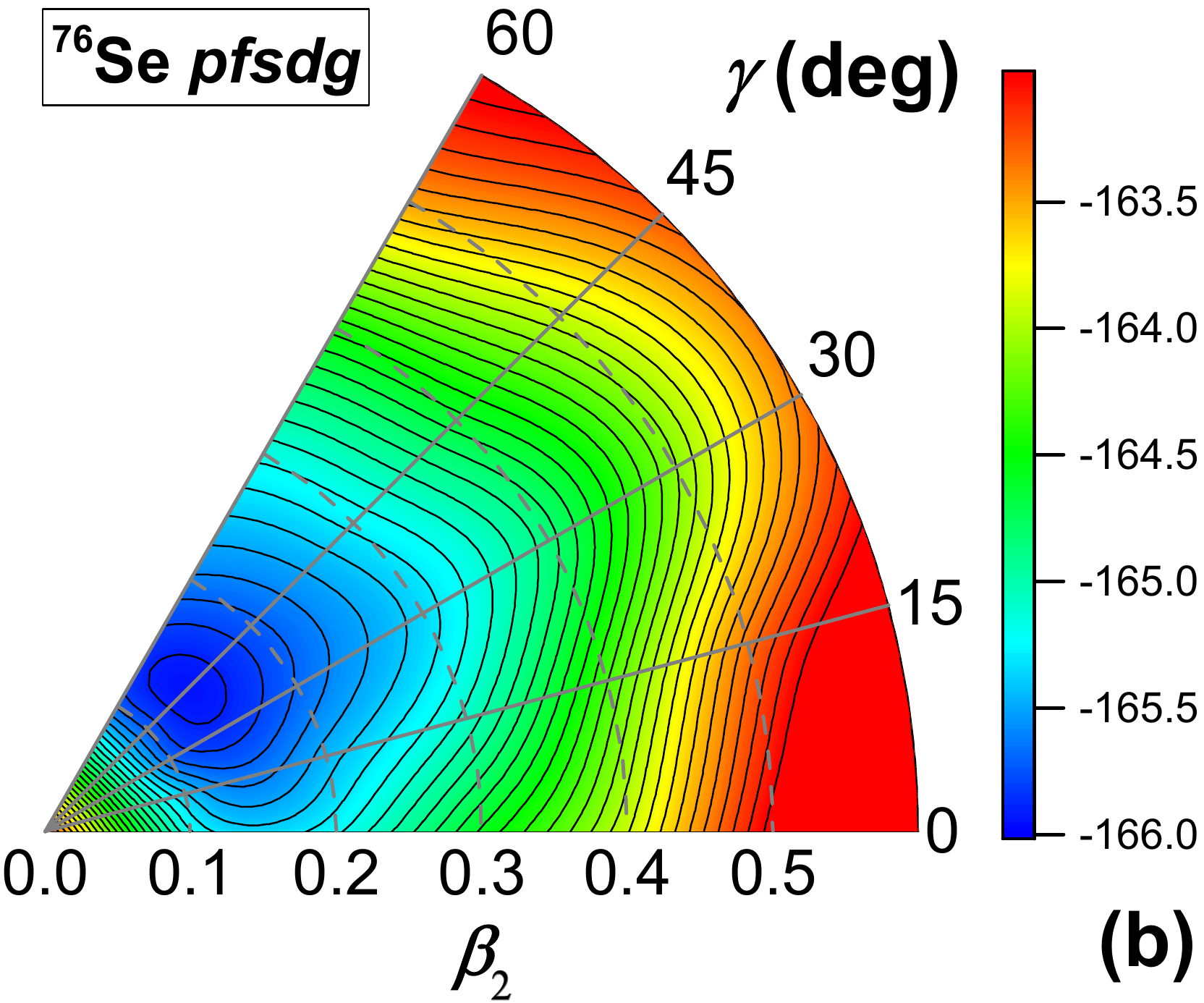}  
\end{minipage}
\caption{\label{fig:pes-2-shell} Potential-energy surfaces for $^{76}$Ge (top)
and $^{76}$Se (bottom) with the Hamiltonian $pfsdg$.}
\end{figure}

\begin{figure}[t]
\includegraphics[width=\columnwidth]{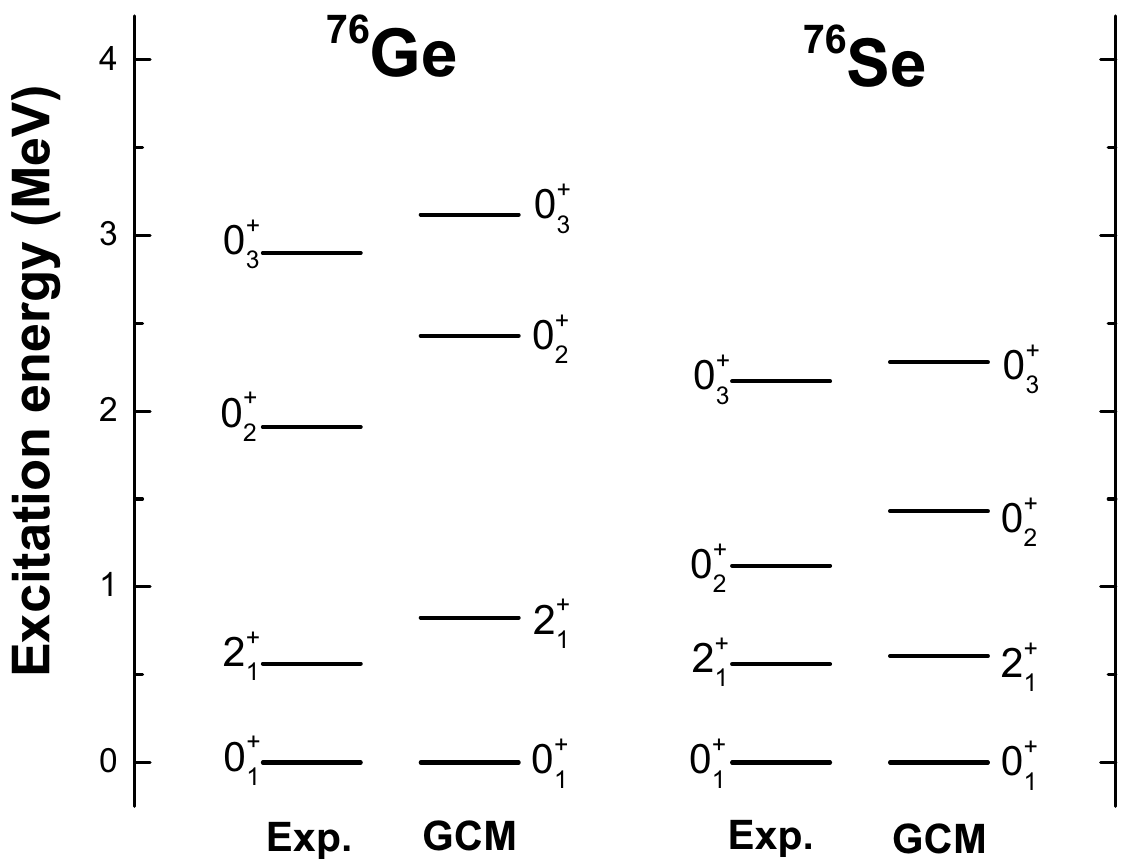}
\caption{\label{fig:spec-two-shell} Calculated low-lying excitation spectra of
$^{76}$Ge and $^{76}$Se produced by the Hamiltonian $pfsdg$ alongside
experimental data~\cite{NNDC}.}
\end{figure}

Figure \ref{fig:pes-2-shell} shows the resulting potential-energy surfaces.  The
minimum occurs at smaller deformation than in the one-shell example so that, as
the calculated spectra in Fig.\ \ref{fig:spec-two-shell} show, the energy of the
first $2^+$ state rises noticeably.  This worsens the agreement with experiment
slightly in $^{76}$Ge, but improves it in $^{76}$Se.  The excited $0^+$ states
are also generally better reproduced in the two-shell calculation.  
 
The $B(E2)$ values calculated in two shells are not better than those from one
shell, however.  With the same effective charges as before, we find noticeably
smaller values: $0.126$ in $^{76}$Ge and $0.221$ in $^{76}$Se.  The
corresponding experimental values are larger, $0.274$ and $0.432$.  Evidently,
the $E2$ operator must be renormalized more in two shells than in one, a result
that is consistent with the smaller deformation in the two-shell calculation but
is nevertheless a little surprising.

We turn finally to the \bbz matrix elements, which appear in Table
\ref{t:two-shell}.  The total matrix element, once triaxial deformation is
included, is only slightly larger than that from GCN2850 in a single shell.
Though our interaction is clearly not perfect, the result suggests that
enlarging the space further may not dramatically change the matrix element,
though gradual but continual changes with the addition of successive shells
cannot be ruled out.  It also shows the importance of including triaxial shapes
in larger spaces.

\begin{table}[b]
\centering
\begin{tabular}{r@{\hskip 1cm}dd}
\toprule &\multicolumn{1}{c}{axial} & \multicolumn{1}{c}{full} \\ \midrule
$M^{0\nu}_{\rm GT}$                  & 3.25       & 2.01   \\
$-\frac{g_V^2}{g_A^2}M^{0\nu}_{\rm F}$ & 0.43       & 0.35   \\ 
$M^{0\nu}_{\rm T}$                   &-0.03       &-0.02   \\ \midrule
Total $M^{0\nu}$                     & 3.65       & 2.34   \\ \bottomrule
\end{tabular}
\caption{\label{t:two-shell}GCM results for thte Gamow-Teller ($M^{0\nu}_{\rm
GT}$), Fermi ($M^{0\nu}_{\rm F}$), and tensor ($M^{0\nu}_{\rm T}$) \bbz matrix
elements for the decay of $^{76}$Ge in two shells, with and without triaxial
deformation.}
\end{table}

\begin{figure}[t]
\includegraphics[width=\columnwidth]{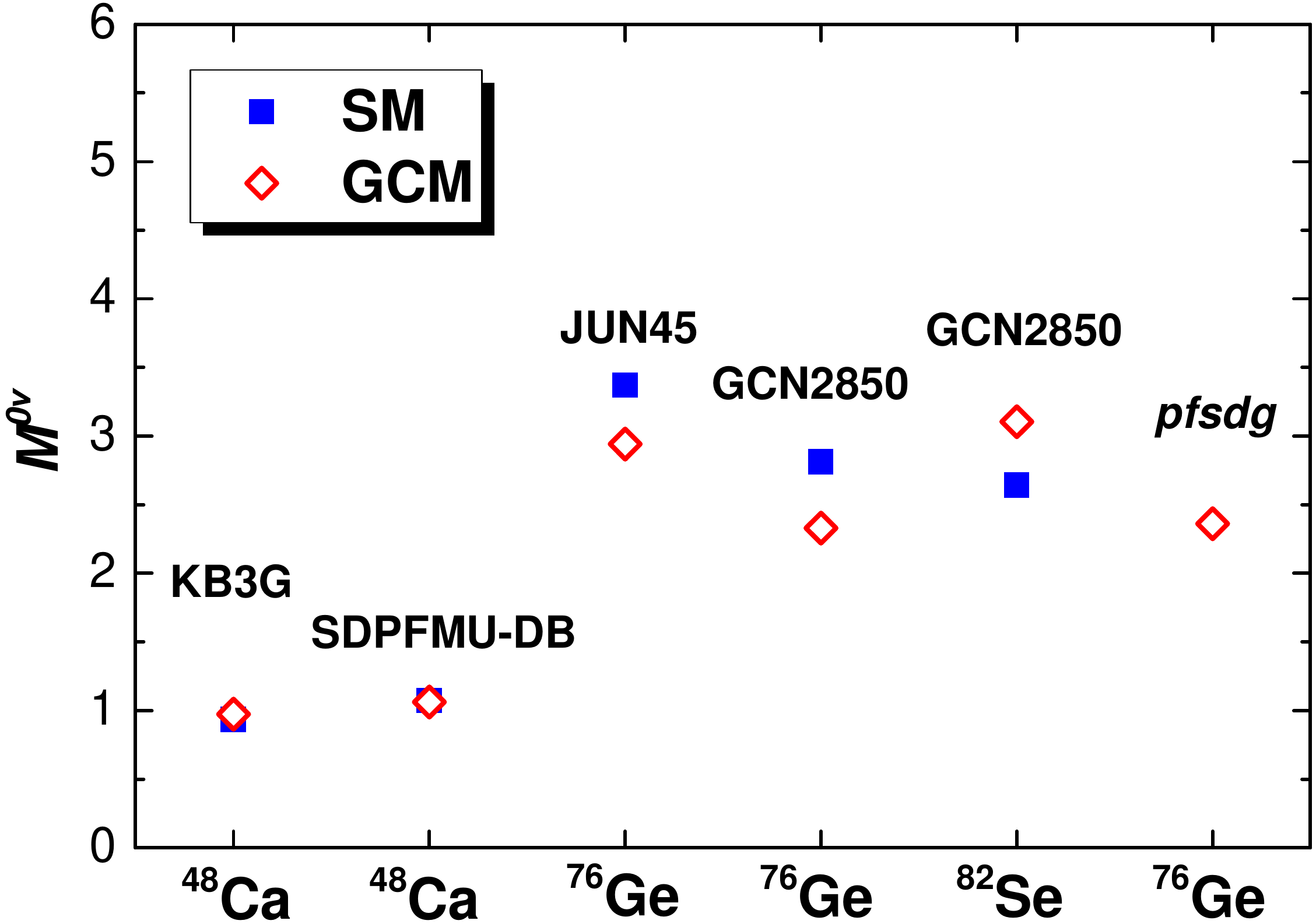}      
\caption{\label{fig:monu} GCM matrix elements $M^{0\nu}$ compared with those of
the shell-model (SM), with either the JUN45~\cite{sen14b}, CN2850~\cite{men08}
KB3G~\cite{men16}, or SDPFMU-DB~\cite{Iwata16} interactions. The term $pfsdg$
denotes the two-shell interaction used here for $A\approx 80$ nuclei.}
\end{figure}

Figure \ref{fig:monu} summarizes our \bb results.  For the decay of $^{48}$Ca,
as noted, we reproduce the exact shell model results nearly perfectly in both
one and two shells.  For the decay of $^{76}$Ge (and $^{82}$Se) in a single
shell, the GCM reproduces the exact result well enough, with two different
effective interactions.  And in two shells, with a brand new effective
interaction, it obtains a result that is only slightly different from the
GCN2850 result in one shell.  

An important caveat, in addition to those already mentioned:  We really ought to
be using an effective \bbz operator to accompany our effective interaction, as
in Refs.\ \cite{hol13c} and \cite{engel09}.  Those papers lead us to suspect a
change of 20\% or less from an effective decay operator in two shells.  In any
event, because we made significant phenomenological adjustments to the
single-particle energies in the prototype calculation here, we cannot
systematically construct the decay operator that should accompany the effective
interaction.

\section{\label{summary}Summary}

The perfect many-body method will include all possible correlations in an
infinitely-large space.  One step on the way to that ideal is to enlarge the
single-particle space for the shell-model, a method that includes all
correlations within that space.  Here we have approximately diagonalized a
shell-model Hamiltonian in two major shells.  Tests in a single shell, and in
two shells for $^{48}$Ca, show that the our approximation method includes the
most important correlations.  Our first-of-its-kind two-shell calculation of the
\bbz matrix element in $^{76}$Ge suggests a small effect from the extra
single-particle orbitals.

There are at least two ways forward from here.  We should use a better effective
Hamiltonian, either by normal-ordering with respect to an ensemble reference
\cite{Stroberg17} that better includes bulk effects of three-nucleon forces far
from closed shells, or by careful tuning of the interaction.  The second option,
besides being very difficult, would make it impossible to develop a consistent
effective operator, but the first should be pursued.  One might also use our GCM
wave functions as a starting point for refinement by the ``multi-reference''
version of the In-Medium Similarity Renormalization Group.  Work in that
direction is in progress.

\section{Acknowledgments}

We would like to thank J. Men\'endez for providing us the unpublished effective
interaction GCN2850, J.\ Simonis for providing matrix elements of normal-ordered
three-nucleon interactions, and N. Hinohara, M.  Horoi, J. Men\'endez, T. R.
Rodr{\'i}guez, and J. M. Yao for helpful discussions.  This work has been
supported by U.S.\ Department of Energy grants DE-FG0297ER41019, DE-SC0008641,
and DE-SC0004142, by by the National Research Council of Canada, and by NSERC.
We used allocations of computing resources at the U.S.\ National Energy Research
Scientific Computing Center (NERSC) and the J\"ulich Supercomputing Center
(JURECA) to carry out computations.  Finally, we thank the Institute for Nuclear
Theory at the University of Washington for its hospitality and the Department of
Energy for partial support during the completion of this work.

%\bibliography{bbchangfeng}


\end{document}